\documentclass[aps,prl,twocolumn,floatfix,a4paper,showpacs,superscriptaddress]{revtex4}
\usepackage{epsfig}
\usepackage{amsmath}
\usepackage{color}

\newcommand{\psinit}{\psi_\mathrm{init}}
\newcommand{\ii}{\mathrm{i}}
\newcommand{\e}{\mathrm{e}}
\newcommand{\op}[1]{\hat{#1}}
\newcommand{\ke}[1]{\left|{#1}\right>}

\newcommand{\Ctrl}{\mathcal C}

\newcommand{\spec}{{\mathcal K}}
\newcommand{\all}{{\mathcal N}}
\newcommand{\subspace}{{\mathcal S}}

\begin{document}
\title{Searching via walking: How to find a marked subgraph of a graph using quantum walks}

\author{Mark Hillery}
\affiliation{Department of Physics, Hunter College of the City University of New York, 695 Park Avenue, New York, NY 10021}
\author{Daniel Reitzner}
\affiliation{Research Center for Quantum Information, Slovak Academy of Sciences, D\'ubravsk\'a cesta 9, 845 11 Bratislava, Slovakia}
\author{Vladim\'\i r Bu\v zek}
\affiliation{Research Center for Quantum Information, Slovak Academy of Sciences, D\'ubravsk\'a cesta 9, 845 11 Bratislava, Slovakia}

\begin{abstract}
We show how a quantum walk can be used to find a marked edge or a marked complete subgraph of a complete graph. We employ a version of a quantum walk, the scattering walk, which lends itself to experimental implementation.  The edges are marked by adding elements to them that impart a specific phase shift to the particle as it enters or leaves the edge. If the complete graph has $N$ vertices and the subgraph has $K$ vertices, the particle becomes localized on the subgraph in $O(N/K)$ steps.  This leads to a quantum search that is quadratically faster than a corresponding classical search.  We show how to implement the quantum walk using a quantum circuit and a quantum oracle, which allows us to specify the resource needed for a quantitative comparison of the efficiency of classical and quantum searches --- the number of oracle calls. \end{abstract}

\maketitle

A quantum walk is a quantum version of a classical random walk. A classical
random walk is described in terms of probabilities; a particle ``sitting'' (localized) on a vertex of a graph
has certain probabilities
to leave that vertex along different edges.  A quantum walk is described in terms of probability amplitudes, and
that means that there are interference effects in quantum walks, which are completely absent in
classical random walks.  The time evolution in these walks can either be in discrete steps \cite{AhDaZa93,AhAmKeVa01} or continuous \cite{FaGu98b}. Both types of quantum walks have proven to be fruitful sources of quantum algorithms \cite{ambainis07,ChEi05,ChClDeFaGuSp03,FaGoGu07}.  A summary of both the properties of quantum walks and their algorithmic applications can be found in a review \cite{kendon07}.

Recently, quantum walks have entered the realm of experiment. Theoretical predictions for the quantum walk on a line have been confirmed in various physical systems --- with an ion in a linear trap performing three ``quantum'' steps in Ref.~\cite{ScMaScGlEnHuSc09}, with neutral atoms in an optical lattice in Ref.~\cite{KaFoChStAlMeWi09} or with photons in waveguide lattices with negligible decoherence in Ref.~\cite{PeLaPoSoMoSi08}. Interesting realizations are also those exploiting the fact, that the most important ingredient in quantum walks is the interference. To this end the wave nature of classical light was exploited in Refs.~\cite{KnRoSi03,JePaKi04} to propose an analog of a quantum walk, and this analog had been experimentally realized in Ref.~\cite{BoMaKaScWo99} though for other
purposes. The use of interferometers to realize quantum walks has also been proposed \cite{JePaKi04}
and also recently accomplished \cite{ScCaPo09}.

Quantum walks on a line have abundance of interesting properties (see e.g. \cite{kendon07}),
but quantum walks, when considered on more complex graphs provide us with deeper understanding of quantum dynamics and provide us
with new insights and hints how to construct new quantum algorithms. In particular, quantum walks have been used to investigate searches on a number of different graphs. In these searches, one of the vertices is distinguished, and the objective is to find that specific (marked) vertex. The graphs considered so far have been grids and hypercubes of different dimensions and the complete graph \cite{ShKeWh03,AmKeRi05,ChGo04}. However, up to now, there has been no experimental realization of a search by means of a quantum walk. One way to achieve this goal may be the adoption of the scattering quantum walk formalism introduced in Ref.~\cite{HiBeFe03} which is based on an interferometric analogy of a scattering process. For this type of a walk, multiport devices are needed. Optically these can be constructed from simpler devices, such as beam splitters, and there is work ongoing to construct multiports for atoms \cite{CoHaDu09}.

Suppose, that instead of finding a distinguished (marked) vertex, one is interested in finding a distinguished edge
or even a distinguished subgraph.  The case of an edge can be viewed as finding two elements in
a list that have a particular relation.  More specifically, let us suppose that $x$ and $y$ are elements of a set,
$\all$, and that $f(x,y)$ is a classical boolean oracle function on $\all\times\all$, such that
\begin{equation}
\label{eq:classical_oracle}
f(k,l)=\begin{cases}
1\text{ if $k\in\spec$ and $l\in\spec$,}\\
0\text{ otherwise,}
\end{cases}
\end{equation}
where $\spec\subset\all$. This function then gives us the answer to the question of whether the two elements, $k$ and $l$, satisfy the relation. Hence, we are interested in finding pairs $(x,y)$ such that $f(x,y)=1$.  There might be a single such pair, which would correspond to finding a single edge, or there might be several.

We shall pursue this study by using the formalism of scattering quantum walks (\cite{HiBeFe03})
in which the particle resides on the edges of a graph rather than on its vertices.  In Ref.~\cite{ReHiFeBu09} we used this formalism to study searches on graphs with high symmetry, including complete graphs and various versions
of multipartite graphs.

Having a graph $\mathcal G=(V,E)$ on which the walk is defined, with $V$ being the set of vertices and $E$ the set of edges, the Hilbert space is defined as
\[
\mathcal H=\ell^2(\{\ke{m,l}|m,l\in V,ml\in E\}).
\]
This definition implies, that the Hilbert space is given by the span of all \emph{edge states}, i.e.~position states $\ke{m,l}$ interpreted as a particle going from vertex $m\in V$ to vertex $l\in V$, with $ml\in E$ being an edge of the graph $\mathcal G$. These edge states form an orthonormal basis of the Hilbert space, which we shall call \emph{the canonical basis.}

In this Hilbert space the unitary evolution is given by a set of \emph{local unitary evolutions} defined for each vertex. If we specify (for every $m\in V$) $A_m=\ell^2(\{\ke{m,l}|l\in V, ml\in E\})$, the set of all edge states originating on vertex $m$, and $\Omega_m=\ell^2(\{\ke{l,m}|l\in V, lm\in E\})$, the set of all edge states ending on vertex $m$, then local unitary evolutions act as $\op U^{(m)}:\Omega_m\to A_m$. The overall unitary step operator, $\op U$, acting on the system is represented by the combined action of the local unitary evolutions, that is, the restriction of $\op U$ to $\Omega_{m}$ is just $\op U^{(m)}$. Given the initial state of the system is $\ke{\psinit}$, the state after $n$ steps is $\ke{\psi_n}=\op U^n\ke\psinit$ and the probability of finding the particle (walker) in state $\ke{k,l}$ is then $|\langle k,l|\psi_n\rangle|^2$.

Let us consider a walk on a complete graph with vertices given by the set $V=\all$ containing $N$ elements.  The initial state is taken to be the equal superposition of all the edge states
\begin{equation}
\label{initstate}
|\psinit \rangle = \frac{1}{\sqrt{N(N-1)}} \sum_{l\in\all}\sum _{\substack{m\in\all \\ m\neq l}} | l,m\rangle .
\end{equation}
This choice of initial state is motivated by the fact that we have no \emph{a priori} knowledge about which edges are marked. The local unitary evolution associated with vertices that are not attached to any of the marked edges is given by
\begin{equation}
\op U^{(l)}\ke{k,l}=-r\ke{l,k}+t\sum_{\substack{m\in\all\\ m\neq l,k}}\ke{l,m},
\label{eq:unitary}
\end{equation}
where $r$ and $t$ are reflection and transmission coefficients whose values are
\[
t=\frac{2}{N-1},\qquad r=1-t.
\]
This choice of local unitary operators for the scattering walk is analogous to the choice of the \emph{Grover coin,} see Ref.~\cite{MaBaStSa02}, in a coined quantum walk.

The target edges are marked by placing ``phase shif\-ters'' on both ends.  These have the effect of modifying the
local unitary operations associated with the vertices to which the edge is attached.  A particle entering or
leaving the edge picks up a phase factor of $e^{i\phi}$, and one that is reflected back into the edge
picks up a factor of $e^{2i\phi}$.  In more detail, if the edge between vertices $j$ and $k$ is the
only marked edge in the graph, we will have
\begin{eqnarray*}
U |j,k\rangle & = & -re^{2i\phi} |k,j\rangle + t e^{i\phi} \sum_{\substack{l\in\all \\ l\neq j,k}}|k,l\rangle, \\
U |m,j\rangle & = & -r |j,m\rangle + te^{i\phi} |j,k\rangle + t \sum_{\substack{l\in\all \\ l\neq j,m,k}} |j,l\rangle.
\end{eqnarray*}

\begin{figure}
\includegraphics[scale=1]{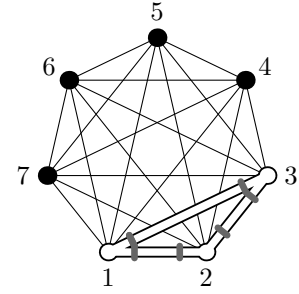}
\caption{\label{fig:CG}An example of a complete graph with $N=7$ vertices out of which $v=3$ are special (white ones). The edge between two such vertices is considered to be special and it performs a phase shift for particle either entering or leaving it (grey bars).}
\end{figure}

Let us now consider a specific type of relation,$R$, between vertices (see 
Eq.~[\ref{eq:classical_oracle})].  We specify a subset of vertices, and two vertices satisfy $R$ if they are both in the specified subset.  These vertices, and the edges
connecting them, will form a complete subgraph.  Consequently, we consider the problem of a scattering quantum walk on a complete graph with a marked complete
subgraph (see Fig.~\ref{fig:CG}).
In particular, let the set of $N$ vertices of the complete graph be $\all$, and the set of $K$
vertices connected by marked edges and forming a complete subgraph be $\spec$.  A quantum
walk on this graph starting from the initial state given in Eq.\ (\ref{initstate}) will take place in a
small subspace of the overall Hilbert space of the walk.  This phenomenon, the reduction of the
effective dimension due to the symmetry of the graph, has been analyzed in detail
for coined walks by Krovi and Brun \cite{KrBr07} and for the scattering walk in Ref.~\cite{ReHiFeBu09}.
We begin by defining four vectors:
\begin{eqnarray*}
\ke{w_1} &=& \frac{1}{\sqrt{K(N-K)}}\sum_{j\in\all\setminus\spec}\sum_{k\in\spec}\ke{j,k},\\
\ke{w_2} &=& \frac{1}{\sqrt{K(N-K)}}\sum_{j\in\spec}\sum_{k\in\all\setminus\spec}\ke{j,k},\\
\ke{w_3} &=& \frac{1}{\sqrt{(N-K)(N-K-1)}}\sum_{j\in\all\setminus\spec} \sum_{\substack{k\in\all\setminus\spec\\k\neq j}}\ke{j,k},\\
\ke{w_4} &=& \frac{1}{\sqrt{K(K-1)}}\sum_{j\in\spec}\sum_{\substack{k\in\spec\\k\neq j}}\ke{j,k}.
\end{eqnarray*}
These vectors form a basis of a subspace of the Hilbert space $\mathcal H$, which we shall denote
by $\subspace$, which is invariant under the action of the unitary operator $\op U$ that advances the walk
one step.  The action of $\op U$ on the basis vectors of $\subspace$ is given by
\begin{eqnarray*}
\op U\ke{w_1} &=& [r-(k-2)t]\ke{w_2}+t\e^{\ii\phi}\sqrt{(k-1)(N-k)}\ke{w_4},\\
\op U\ke{w_2} &=& [(k-1)t-r]\ke{w_1}+t\sqrt{k(N-k-1)}\ke{w_3},\\
\op U\ke{w_3} &=& t\sqrt{k(N-k-1)}\ke{w_1}+[r-t(k-1)]\ke{w_3},\\
\op U\ke{w_4} &=& t\e^{\ii\phi}\sqrt{(k-1)(N-k)}\ke{w_2}\\
& & +[t(k-2)-r]\e^{2\ii\phi}\ke{w_4}.
\end{eqnarray*}
For the evolution to remain entirely within $\subspace$ the initial state must be in this subspace as well. Our
initial state [see Eq.~(\ref{initstate})] can be expressed as
\begin{eqnarray*}
\ke\psinit & = & \sqrt{\frac{k(N-k)}{N(N-1)}}(\ke{w_1}+\ke{w_2})+\sqrt{\frac{k(k-1)}{N(N-1)}}\ke{w_4} \\
& & +\sqrt{\frac{(N-k)(N-k-1)}{N(N-1)}}\ke{w_3} ,
\end{eqnarray*}
and is, therefore in $\subspace$.  Consequently, in order to determine how this particular walk evolves, we
need to only consider a four-dimenional problem.  One simply finds the eigenvalues, $\lambda_{\mu}$
and eigenstates $|\mu\rangle$, where $\mu =1, \ldots, 4$, of $\op U$ restricted to $\subspace$, and
finds the state after $n$ steps by exploiting expansion
\begin{eqnarray*}
\ke{\psi_n} & = & \op U^n\ke\psinit=\op U^n\sum_\mu\left<\mu|\psinit\right>\ke\mu\\
& = & \sum_\mu\lambda_\mu^n\left<\mu|\psinit\right>\ke\mu.
\end{eqnarray*}

We now need to specify the value of $\phi$.  For localizing the particle on the marked edges, we employ phase-shift $\phi =\pi /2$.  In the case  $N\gg k$ the expression for the vector after $n$ steps reads
\[
\ke{\psi_n}\simeq\begin{pmatrix}
0\\ 0\\ \cos 2xn\\ \ii\sin 2xn
\end{pmatrix},
\]
where the first entry is the $|w_{1}\rangle$ component, the second is the $|w_{2}\rangle$ component,
and so on.  In this expression,
\[
x=\frac{\sqrt{K(K-1)}}{N-1} ,
\]
and the terms that have been neglected are $O(\sqrt{x})$ or smaller.  We see that after
$n=\pi /(4x) = O(N/K)$ steps, the particle is in the state $|w_{4}\rangle$, which means it is localized
on the marked edges.  After running the walk the proper number of steps to localize the particle
on the marked edges, we complete the search by measuring the position of the walker to see on which edge it is located.

If we are searching for a single edge, i.e.\ $K=2$, then the quantum search represents a quadratic
speedup over what is possible classically. Classically we would just check each edge to see whether
it is marked or not, and we would have to check $O(N^{2})$ edges in order to find the marked one
(this will be made more precise shortly).  This quantum advantage remains for small subgraphs.  For example,
if the subgraph is a triangle, our probability of finding all three vertices after running the quantum
search twice is $2/3$ and the probability of finding all three vertices after no more than three runs of the search is
$8/9$.  The expectation value of the number of searches necessary to find all three vertices
is $5/2$.  Things become more
complicated if the subgraph is a complete graph on four vertices, because there are more alternatives.
After two runs of the search, the probability that we have found all four vertices is $1/6$ and the
probability that we have found three out of four of them is $2/3$.  If we go to three runs, the probability
of finding all four vertices becomes $19/36$ and the probability of finding three out of four becomes
$4/9$.  Therefore, for small subgraphs, a small number of runs of the walk will allow us to find all
vertices of the subgraph with high probability.

To provide a different perspective on the proposed evolution unitary $\op U$, we rephrase the quantum walk in terms of a quantum circuit in which the procedure of checking an
edge to see whether it is marked is a call to a quantum oracle, corresponding to the classical oracle given by Eq.~(\ref{eq:classical_oracle}). The quantum oracle can be interpreted as a unitary operation acting on a tripartite system as
\begin{equation}
\label{eq:oracle}
\mathcal C\op U_f\ke k\otimes\ke l\otimes\ke m=\ke k\otimes\ke l\otimes\ke{m\oplus_4 f(k,l)}
\end{equation}
for $k,l\in\all$, where the first two subsystems are both $N$ dimensional, the last one is from a four-dimensional Hilbert space and $\oplus_4$ is the addition modulo four. In this formulation, it becomes clearer what  resources are being compared in the quantum and classical cases --- the
particular resource we are focussing on is the number of oracle calls.
To find one pair marked by the oracle from Eq.~(\ref{eq:oracle}), we would classically (when we are not allowed to use the interference) need to query the oracle $O[(N/K)^2]$ times. In the quantum case, however, we have seen that only $O(N/K)$ queries are needed to find the pair with high probability,
because that is the number of steps a quantum walk would need to localize the particle on the
marked edge, and, as we shall see, the oracle is called only twice per step.

\begin{figure}
\includegraphics[scale=1]{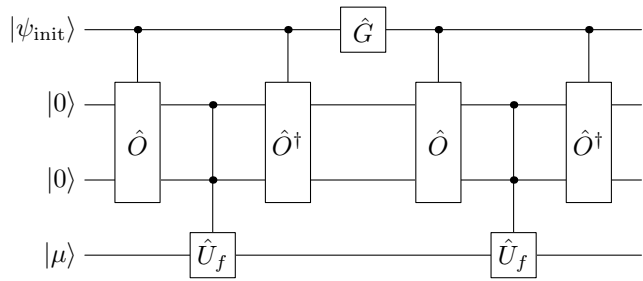}
\caption{\label{fig:circuit}
A quantum circuit (network) that implements a single step of scattering quantum walk search, which makes use of the quantum oracle $\Ctrl\op U_f$. The first input corresponds to a quantum walker originally prepared in the state $\ke\psinit$. The second and third inputs represent the end vertices of the walker's edge state, while the fourth input represents an ancillary processing subsystem prepared in the state $\ke\mu$ given by Eq.~(\ref{eq:oracle_eigenstate}).}
\end{figure}

The oracle is incorporated into the scattering quantum walk as shown in Fig.~\ref{fig:circuit}, where in addition to the walking Hilbert space we make use of the three ancillary systems that the quantum oracle acts upon. To obtain the information about the actual vertices we use a unitary gate $\op O$, whose action can be expressed as the action of two controlled operations similar to those given in Ref.~\cite{ReHiFeBu09}. On the state $\ke{k,l}\otimes\ke 0\otimes\ke 0\otimes\ke m$, $\op O$ acts as
\[
\op O\ke{k,l}\otimes\ke 0\otimes\ke 0\otimes\ke m=\ke{k,l}\otimes\ke k\otimes\ke l\otimes\ke m.
\]

It is useful to prepare the last subsystem in a special state
\begin{equation}
\label{eq:oracle_eigenstate}
\ke\mu=\frac{1}{2}\sum_{q=0}^3\e^{-\frac{\ii\pi q}{2}}\ke q.
\end{equation}
The usefulness can be seen from equality
\begin{multline*}
\Ctrl\op U_f\ke{k,l}\otimes\ke k\otimes\ke l\otimes\ke\mu=\\
=\e^{\ii\frac{\pi}{2} f(k,l)}\ke{k,l}\otimes\ke k\otimes\ke l\otimes\ke\mu.
\end{multline*}
So we see, that the composed operation $\op O^\dagger\Ctrl\op U_f\op O$ changes the state $\ke{k,l}\otimes\ke 0\otimes\ke 0\otimes\ke\mu$ to the state $\e^{\ii\frac{\pi}{2} f(k,l)}\ke{k,l}\otimes\ke 0\otimes\ke 0\otimes\ke\mu$. In this way the ancillary systems assist the evolution, but the ancillary systems do not themselves change, making it unnecessary to express them every time. This means, that this view is isomorphic to the one where we did not make use of the quantum oracle. As each step of the walk involves only two oracle calls, and with the number of oracle calls being a measure of the complexity of the problem, we conclude that fewer oracle calls are needed in the quantum case than in the classical for $N\gg k\geq 1$.

In Ref.~\cite{ReHiFeBu09} we considered scattering-quantum-walk searches on several examples of highly symmetric (complete, bipartite, and $M$-partite) graphs where some of the vertices were special.  In particular, the special vertices simply reflected the particle with a phase factor of $\exp (i\phi)$. As is the case here, the symmetry of these graphs led to a significant reduction in the dimensionality of the problem. For all of the types of graphs we considered, we found a quadratic speedup over the classical search when the phase-shift of special vertices was taken to be $\pi$. We see from our analysis here that if one wants to find an edge, a different phase shift is required.

In conclusion, we introduced a novel application of quantum walks. Specifically, quantum walks are used to find a marked edge, or a marked subgraph, in a complete graph.  We proved that the quantum walk can perform the search (quadratically) faster than it is possible classically. One of the attractive features of our model is that it might be straightforwardly realized in simple scattering experiments.

\section*{Acknowledgements}
Our work has been supported by projects QAP 2004-IST-FETPI-15848, HIP FP7-ICT-2007-C-221889, APVV QIAM, CE SAV QUTE and by the National Science Foundation under grant number PHY-0903660.


\begin{thebibliography}{10}

\bibitem{AhDaZa93}
Y.~Aharonov, L.~Davidovich, and N.~Zagury,
\newblock {\em Phys. Rev. A}{ \bf 48}(2), 1687 (1993).

\bibitem{AhAmKeVa01}
D.~Aharonov, A.~Ambainis, J.~Kempe, and U.~Vazirani,
\newblock In {\em Proc. of the 33rd ACM STOC},  50--59 (2001).

\bibitem{FaGu98b}
E.~Farhi, and S.~Gutmann,
\newblock {\em Phys. Rev. A}{ \bf 58}(2), 915 (1998).

\bibitem{ambainis07}
A.~Ambainis,
\newblock {\em SIAM J. on Comp.}{ \bf 37}(1), 210 (2007).

\bibitem{ChEi05}
A.~Childs, and J.~Eisenberg,
\newblock {\em Quantum Information and Computation}{ \bf 5}(7), 593 (2005).

\bibitem{ChClDeFaGuSp03}
A.~Childs, R.~Cleve, E.~Deotto, E.~Farhi, S.~Gutmann, and D.~Spielman,
\newblock In {\em Proc. of the 35th ACM STOC},  59--68 (2003).

\bibitem{FaGoGu07}
E.~Farhi, J.~Goldstone, and S.~Gutmann,
\newblock \emph{Preprint} arXiv:quant-ph/0702144, (2007).

\bibitem{kendon07}
V.~Kendon,
\newblock {\em Math. Struct. in Comp. Sci.}{ \bf 17}(06), 1169 (2007).

\bibitem{ScMaScGlEnHuSc09}
H.~Schmitz, R.~Matjeschk, C.~Schneider, J.~Glueckert, M.~Enderlein, T.~Huber,
  and T.~Schaetz,
\newblock {\em Phys. Rev. Lett.}{ \bf 103}(9), 090504 (2009).

\bibitem{KaFoChStAlMeWi09}
M.~Karski, L.~Forster, J.-M.~Choi, A.~Steffen, W.~Alt, D.~Meschede, and
  A.~Widera,
\newblock {\em Science}{ \bf 325}(5937), 174--177 (2009).

\bibitem{PeLaPoSoMoSi08}
H.B.~Perets, Y.~Lahini, F.~Pozzi, M.~Sorel, R.~Morandotti, and
  Y.~Silberberg,
\newblock {\em Phys. Rev. Lett.}{ \bf 100}(17), 170506 (2008).

\bibitem{KnRoSi03}
P.L.~Knight, E.~Rold\'an, J.E.~and Sipe,
\newblock {\em Phys. Rev. A}{ \bf 68}(2), 020301(R) (2003).

\bibitem{BoMaKaScWo99}
D.~Bouwmeester, I.~Marzoli, G.P.~Karman, W.~Schleich, and J.P.~Woerdman,
\newblock {\em Phys. Rev. A}{ \bf 61}(1), 013410 (1999).

\bibitem{JePaKi04}
H.~Jeong, M.~Paternostro, and M.S.~Kim,
\newblock {\em Phys. Rev. A}{ \bf 69}(1), 012310 (2004).

\bibitem{ScCaPo09} 
A.~Schreiber, K.~N.~Cassemiro, V.~Poto\v{c}ek, A.~Gabris, P.~Mosley, E.~Andersson,
I.~Jex, and Ch.~Silberhorn, quant-ph/0910.2197.

\bibitem{ShKeWh03}
N.~Shenvi, J.~Kempe, and Birgitta K.~Whaley,
\newblock {\em Phys. Rev. A}{ \bf 67}(5), 052307 (2003).

\bibitem{AmKeRi05}
A.~Ambainis, J.~Kempe, and A.~Rivosh,
\newblock In {\em Proc. of the 16th ACM-SIAM SODA},  1099--1108 (2005).

\bibitem{ChGo04}
A.M.~Childs, and J.~Goldstone,
\newblock {\em Phys. Rev. A}{ \bf 70}(2), 022314 (2004).

\bibitem{HiBeFe03}
M.~Hillery, J.~Bergou, and E.~Feldman,
\newblock {\em Phys. Rev. A}{ \bf 68}(3), 032314 (2003).

\bibitem{CoHaDu09}
J.~Cooper, D.~Hallwood, and J.~Dunningham,
\newblock {\em Journal of Physics B}{ \bf 42}(10), 105301 (2009).

\bibitem{ReHiFeBu09}
D.~Reitzner, M.~Hillery, E.~Feldman, and V.~Bu\v{z}ek,
\newblock {\em Physical Review A}{ \bf 79}(1), 012323 (2009).

\bibitem{MaBaStSa02}
T.~Mackay, S.~Bartlett, L.~Stephenson, and B.~Sanders,
\newblock {\em Journal of Physics A}{ \bf 35}(12), 2745 (2002).

\bibitem{KrBr07}
H.~Krovi, and T.A.~Brun,
\newblock {\em Physical Review A}{ \bf 75}, 062332 (2007).

\end{thebibliography}

\end{document}